\documentstyle[psfig,twocolumn,aps]{revtex} \tighten
\begin{document}
\draft

\title{\bf On the Incommensurate Phase in Modulated Heisenberg Chains}

\author{F. Sch\"onfeld, G. Bouzerar, G.S. Uhrig and E.
  M\"uller-Hartmann}

\address{ Institut f\"ur Theoretische Physik, Universit\"at zu
  K\"oln,\\ Z\"ulpicher Str. 77, K\"oln 50937, Germany.  \\ 
  {\rm(\today)} }

\address{~ \parbox{14cm}{\rm \medskip Using the density matrix
    renormalization group method (DMRG) we calculate the magnetization
    of frustrated $S=\frac{1}{2}$ Heisenberg chains 
    for various modulation patterns of the nearest neighbour
    coupling: commensurate, incommensurate with sinusoidal modulation
    and incommensurate with solitonic modulation. We focus on the
    order of the phase transition from the commensurate dimerized
    phase (D) to the incommensurate phase (I). It is shown that the
    order of the phase transition depends sensitively on the model. For
    the solitonic model in particular, a $k$-dependent elastic energy
    modifies the order of the transition. Furthermore, we calculate
    gaps in the incommensurate phase in adiabatic approximation.}}

\maketitle


\narrowtext
\section{Introduction}
Low dimensional magnetic systems have attracted considerable attention in
recent years. Various theoretical and experimental efforts have been
made to understand the fascinating low energy physics of quasi-one
dimensional gapped spin systems, such as spin-Peierls systems (
CuGeO$_{3}$\cite{hase93a} or NaV$_2$O$_5$\cite{isobe96,weide97}),
Haldane systems (e.g.
Ni(C$_3$H$_{10}$N$_2$)$_2$N)$_2$ClO$_4$\cite{sieli95}) and spin
ladders (e.g. SrCu$_2$O$_3$ \cite{dagot96} or
Cu$_2$(C$_5$H$_{12}$N$_2$)$_2$Cl$_4$\cite{chabo97}).  Even if 
questions still remain open, many of the experimentally observed features
can be already understood within the framework of one dimensional
Heisenberg chains with various couplings (this includes also spin
ladders \cite{brehm96}).

Some of these systems exhibit interesting features in external
magnetic fields, for instance, a transition from a commensurate to an
incommensurate phase.  At this transition weak hysteresis effects are
observed in CuGeO$_{3}$ at low temperatures \cite{hase93b,loosd96a}
and Kiryukhin {\it et al.} found a small jump in the incommensurability
measured by X-ray scattering \cite{kiryu95,kiryu96a,kiryu96b}.
These features are characteristic for a first order phase transition.
From the theoretical point of view, no consensus has been reached so
far on the order of the transition.  The phase transition was
predicted to be of first order by Cross \cite{cross79b}. Bhattacharjee
{\it et al.} obtained the same conclusion using a phenomenological
Landau expansion \cite{bhatt98}.  But mean-field calculations of
Fujita and Machida for a renormalized XY-model display a second order
phase transition \cite{fujit84} while Buzdin {\it et al.} \cite{buzdi83b} 
find a second order phase transition only at $T=0$ and a first order one 
for $T>0$ using essentially the same model as Fujita/Machida.
Horovitz underlines the importance of the correct treatment of cutoffs
when passing to the continuum limit \cite{horov81a,horov87}.

In this paper, we propose to clarify which parameters influence the
properties of the I phase with the help of the DMRG method for finite
systems.  In section II we calculate magnetizations for different
types of modulations and show that the order of the D--I phase
transition is model dependent. In the I phase we calculate the
magnetization dependence of the two gaps $\Delta_{+/-}$ corresponding
to the increase/decrease of the $z$ component of the total spin 
by unity \cite{uhrig98a}.

For all calculations we have chosen parameter sets which are
convenient for the numerical calculations, i.e. displaying small
finite size effects.  Computational aspects are given in section III.
In section IV we summarize the results.

\section{Magnetization}
In the adiabatic approximation for the phonons the modulation of the
exchange couplings can be described by parameters $\delta_i$ which are
linked to the lattice distortion. Thus the Hamiltonian includes an
elastic energy which is a positive quadratic form of the
$\{\delta_i\}$.  In a first step we take the elastic energy to be
dispersionless, i.e. diagonal in real space
\begin{eqnarray}\label{Hamilton}
\hat H & = & \hat H_{\rm chain} + \hat H_{\rm Zeeman} + 
E_{\rm  elast}
\\ 
\hat H_{\rm chain} & = & \sum_{i=1}^{L} \left( J(i)
  {\bf S}_i \cdot {\bf S}_{i+1} +J \alpha \, {\bf S}_i \cdot {\bf S}_{i+2}
\right) \nonumber 
\\ 
\hat H_{\rm Zeeman} & = & g\mu_{\rm B} H S_z \nonumber \ ,
\\
E_{\rm elast} & = & \frac{K_0}{2} \sum_i \delta_i^2 \nonumber \ , 
\\ 
J(i) & =& J(1+\delta_i) \nonumber \ ,
\end{eqnarray}
where $\alpha$ denotes the relative frustration and $S_z$ is the $z$
component of the total spin of the $L$-site chain. The last two terms
in (\ref{Hamilton}) are the Zeeman energy and the elastic energy 
associated to the lattice distortion.

\subsection{Fixed Modulations}

\begin{figure}
\center{\psfig{width=\columnwidth,height=7cm,file=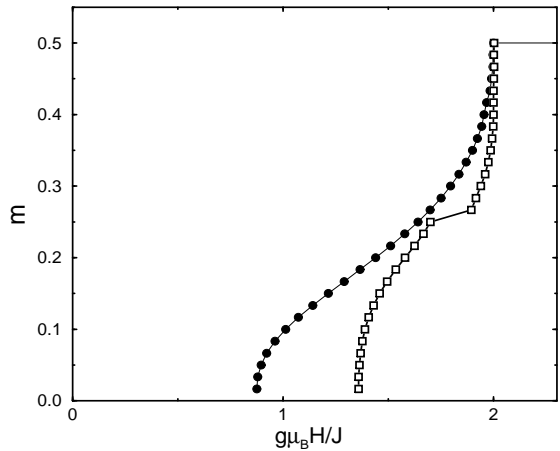}}
\caption[]{Magnetization as a function of the magnetic field for
  $\delta = 0.3$, $\alpha=0.1$ (filled circles) and
  $\delta=0.5$, $\alpha=0.2$ (open squares) for a 60 site chain.}
\label{fig1}
\end{figure}

As a starting point let us consider the simple case where the
lattice distortion is kept frozen as in the D phase even in the
presence of a magnetic field. 
\setcounter{equation}{0}
\renewcommand{\theequation}{\roman{equation}}
\begin{equation}
\label{comm}
\delta_i = (-1)^i \delta \nonumber
\end{equation}
The amplitude $\delta$ is treated as a fixed parameter. The constant 
elastic energy is not taken into account for the moment.  Chitra and 
Giamarchi \cite{chitr97} calculated the magnetization of frustrated {\it or}
dimerized spin chains in a magnetic field using bosonization
techniques. Within this continuum-limit approach the frustration and
dimerization cannot be treated simultaneously (double sine-Gordon
model). For $\alpha < \alpha_{c}$ and $\delta>0$ the
contribution of the frustration is assumed to be irrelevant and the model
reduces to an integrable sine-Gordon model. However, recently it has been 
shown by Bouzerar {\it et al.} that this is not the case 
\cite{bouzerar}. The magnetization
increases just above the lower critical field $H_c$ like $m \propto
\sqrt{H - H_c}$ \cite{chitr97,sakai}. The same power law is found for $\alpha >
\alpha_{c}$ and $\delta=0$. In particular, the transition to finite
magnetization is of {\em second order} in both cases.  With the DMRG
method, we find this square root behavior in presence of both
dimerization and frustration. But as shown 
recently by Tonegawa {\it et al.} by means of exact diagonalization 
\cite{toneg98} an additional remarkable difference appears in some 
parameter range of dimerization and frustration, i.e. a plateau at 
$m=1/4$, see Fig. \ref{fig1}.

In the case of finite magnetization, it is known that the commensurate
dimerization pattern (\ref{comm}) is not appropriate for describing
spin-Peierls systems. For instance X-ray measurements on CuGeO$_3$
clearly show that the structure of the lattice distortion becomes
incommensurate under a sufficiently large magnetic field
\cite{kiryu95,kiryu96a,kiryu96b}.  Thus a more appropriate
choice for the modulation is, 
\begin{equation}
\label{incomm}
\delta_i = \delta \cos (q r_i)
\end{equation}
as it was suggested in \cite{uhrig98a,riera96}. To begin with, 
$q$ is considered as a free parameter which is fixed by minimizing 
the total (free) energy. Note that for $q \neq \pi$ the elastic 
energy is $q$ independent for the ansatz (\ref{incomm}) yielding 
only a constant contribution at given amplitude which will be dropped 
for the following consideration.

\begin{figure}
\center{\psfig{width=\columnwidth,height=7cm,file=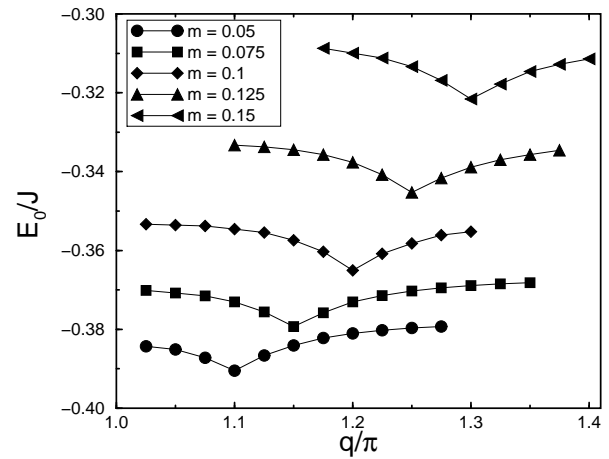}}
\caption[]{Ground state energy per site of an 80 site chain as a function of 
  the wave vector $q$ for various magnetizations $m$, $\alpha = 0.35$ and 
  $\delta =0.1$.}
\label{fig2}
\end{figure}

Using the Jordan-Wigner transformation the applied magnetic field
corresponds to a shift of the chemical potential. For the $XY-$model
with a finite magnetization $m=S_z/L$, it is straightforward to
show that an infinitesimal spin-lattice coupling leads to an
instability at momentum $q = 2k_{\rm F} = \pi(1 + 2m)$.  In the case of the
Heisenberg model, this relation is expected to hold true as well 
\cite{cross79b,uhrig98a,chitr97,mulle81,poilb97}.  We have confirmed 
numerically for various sets of parameters $\delta$, $\alpha$ and
various system sizes that the energy is minimum at $q = \pi(1 + 2m)$
for a given $m$.  The ground state energy per site for $ H=0$ as a function of
$q$ is plotted in Fig.\ref{fig2}, for various magnetizations $m$ at
fixed $\delta =0.1$ and $\alpha = 0.35$. The positions of the cusps
correspond exactly to $q = \pi(1 + 2m)$. The fact that one observes
cusps and not smooth quadratic minima is linked to the divergence of
the susceptibility at $q = \pi(1 + 2m)$, i.e.\ an instability even for
infinitesimal coupling.  This is the generic behavior, independent of
the parameters $\alpha$ and $\delta$, and confirms the relation between
the wave vector and the magnetization.

\begin{figure}
\begin{picture}(9,13)(0,0)
\put(-0.4,4){\psfig{width=\columnwidth,height=8cm,file=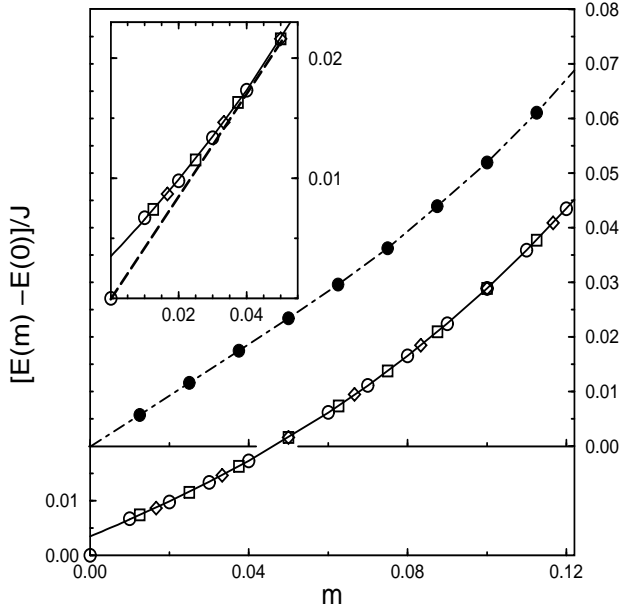}}
\put(-0.3,0){\caption[]{Open symbols (left scale): ground state energies 
  $E(m)-E(0)$ as a function of the 
  magnetization for the sinusoidal modulation (\ref{incomm}) ($\delta =0.2$, 
  $\alpha= 0.35 $) for chains of 100 (circles) 80 (squares) and 60 (diamonds)
  sites. To highlight the discontinuity
  at $m=0$ a cubic fit for $m>0$ is depicted with a solid line. The inset
  shows an enlargement and the tangent for $m=0.05$ as described in the text.
  \newline
  Filled circles (right scale): ground state energies $E(m)-E(0)$ for the adaptive
  modulation 
  from (\ref{deltai}), $K_0 = 1.7$ ($\delta \approx 0.2$ in the D phase) and 
  $\alpha= 0.35$. 
  The dot-dashed line is just a guide to the eye.}
\label{fig3}}
\end{picture}
\end{figure}

Henceforth, we fix $q =\pi(1 + 2m) $ and investigate the magnetization 
as a function of the applied field. We find that the incommensurate exchange
coupling (\ref{incomm}) has a rather strong effect on the magnetization 
leading to a {\it first order} phase transition. To elucidate this we present
 the magnetic ground state energy per site 
$E(m) = (\langle \hat H_{\rm chain}\rangle + E_{\rm elast})/L$ 
as a function of the magnetization in Fig.\ref{fig3}. Results 
for several chain lengths are included to show the absence of finite 
size effects. 

The salient feature of $E(m)$ for sinusoidal modulation is the
discontinuous jump at $m=0$. To understand this jump it is helpful
to look at the averaged squared distortion $\frac{1}{L}\sum_i \delta_i^2$ 
which takes the value $\delta^2$ at $q=\pi$ and $\delta^2/2$ otherwise. 
We see that in the D phase all $\delta_i$ are maximally distorted whereas in
the sinusoidally modulated phase there are also large regions with
weaker distortion. So neither the elastic energy is not continuous in 
the limit $q\to\pi$ nor is the magnetic energy since it 
reacts to the distortions.  
\unitlength1cm

\begin{figure}
\center{\psfig{width=\columnwidth,height=7cm,file=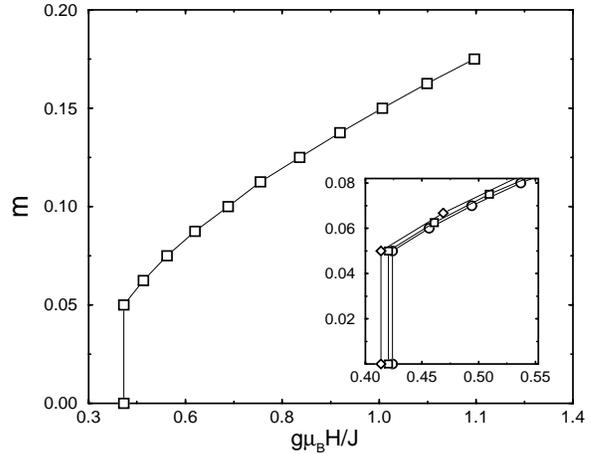}}
\caption[]{Magnetization as a function of the applied magnetic field
  for $\alpha=0.35$ and $\delta = 0.2$ of a 80 site chain as deduced
  from the open squares in Fig.{\protect\ref{fig3}}.
  The inset shows an enlargement near $H_c$ for 100 (circles) 80 
  (squares) and 60 (diamonds) site chains.}
\label{fig4}
\end{figure}

To deduce the dependence $m(H)$ from Fig.\ref{fig3} we have to resort to
Maxwell's construction, i.e.\ we compute the convex hull. The magnetic
field defines the slope $g\mu_{\rm B}H = \partial E/\partial m$ of the
tangent which touches the convex hull at the value $m$ (Legendre
transformation). So one obtains $m(H)$. The jump in
$E(m)$ leads to a first order transition with a jump in $m(H)$. The
resulting $m(H)$ deduced from Fig.\ref{fig3} is depicted in
Fig.\ref{fig4}.

Calculating the corresponding local magnetizations \cite{uhrig98a} 
one finds that there is a large alternating local magnetization close to 
each zero of the modulation. Summing the local magnetizations around each zero
one finds a contribution of $S_z=1/2$, i.e.\ of one spinon.

\subsection{Adaptive Modulations}

\begin{figure}
\begin{picture}(9,8)(0,0)
  \put(0,4.5){$\delta_i$} \put(4.5,1.7){\it i} 
  \put(6.3,6.62){\footnotesize iteration}
  \put(2.4,6.92){\footnotesize sinusoidal}	
  \put(-0.2,1.0){\psfig{width=\columnwidth,height=7cm,file=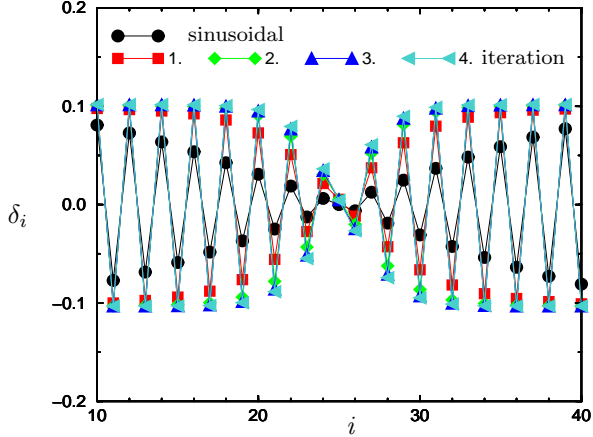}}
  \put(-0.3,0){\caption[]{Incommensurate modulation: local distortions 
  $\delta_i$ versus site index $i$ for the first four iterations starting 
  from a sinusoidal modulation (filled circles); 100 sites, $\alpha = 0.35 $,
  $K_0=3.3$, and $S_z=1$.}
\label{fig5}}
\end{picture}
\end{figure}

In the previous section we chose a sinusoidal modulation and found
that $q = \pi(1 + 2m)$ minimizes the total energy.  
We now proceed in a more general way by minimizing the total energy
including the elastic energy term with respect to all the parameters
$\delta_i$.  In other words, we allow the lattice distortion to adapt
to the spin system. 
Within our DMRG approach we
follow the iterative procedure proposed by Feiguin {\it et al.}
\cite{feigu97} who applied exact diagonalization and Monte-Carlo-Simulations 
to a slightly different model.  The dimerization
amplitudes $\delta_i$ are calculated self-consistently by minimizing
$\langle \hat H_{\rm chain}\rangle + E_{\rm elast}$ which leads to
\begin{equation}
  \label{deltai}
  J\langle {\bf S}_i\cdot{\bf S}_{i+1}\rangle + K_0\delta_i - 
  \frac{J}{L}\sum_{i}
  \langle {\bf S}_i\cdot{\bf S}_{i+1}\rangle = 0 \; ,
\end{equation}
where the last term ensures that the $\delta_i$ satisfy the constraint
$\sum_{i=1}^{L}\delta_i =0$.  Following \cite{feigu97} this equation
is used to improve iteratively the local distortions $\delta_i$. The
expectation values are taken with respect to the ground state of the
previous iteration. Starting from the sinusoidal modulation
(\ref{incomm}) we find that four to five iterations are enough
to achieve a stable pattern that does not change significantly 
on further iterations as shown in Fig.\ref{fig5}.
\unitlength1cm
The envelope of the final modulation can be fitted by a
product of complete Jacobi elliptic functions of modulus $k$ as
predicted analytically\cite{horov81a,buzdi83b,fujit84}. For very low
magnetization, i.e.\ low concentration of solitons, the vicinity of
each zero resembles a $\tanh$ \cite{nakan81}.

\begin{figure}
\center{\psfig{width=\columnwidth,file=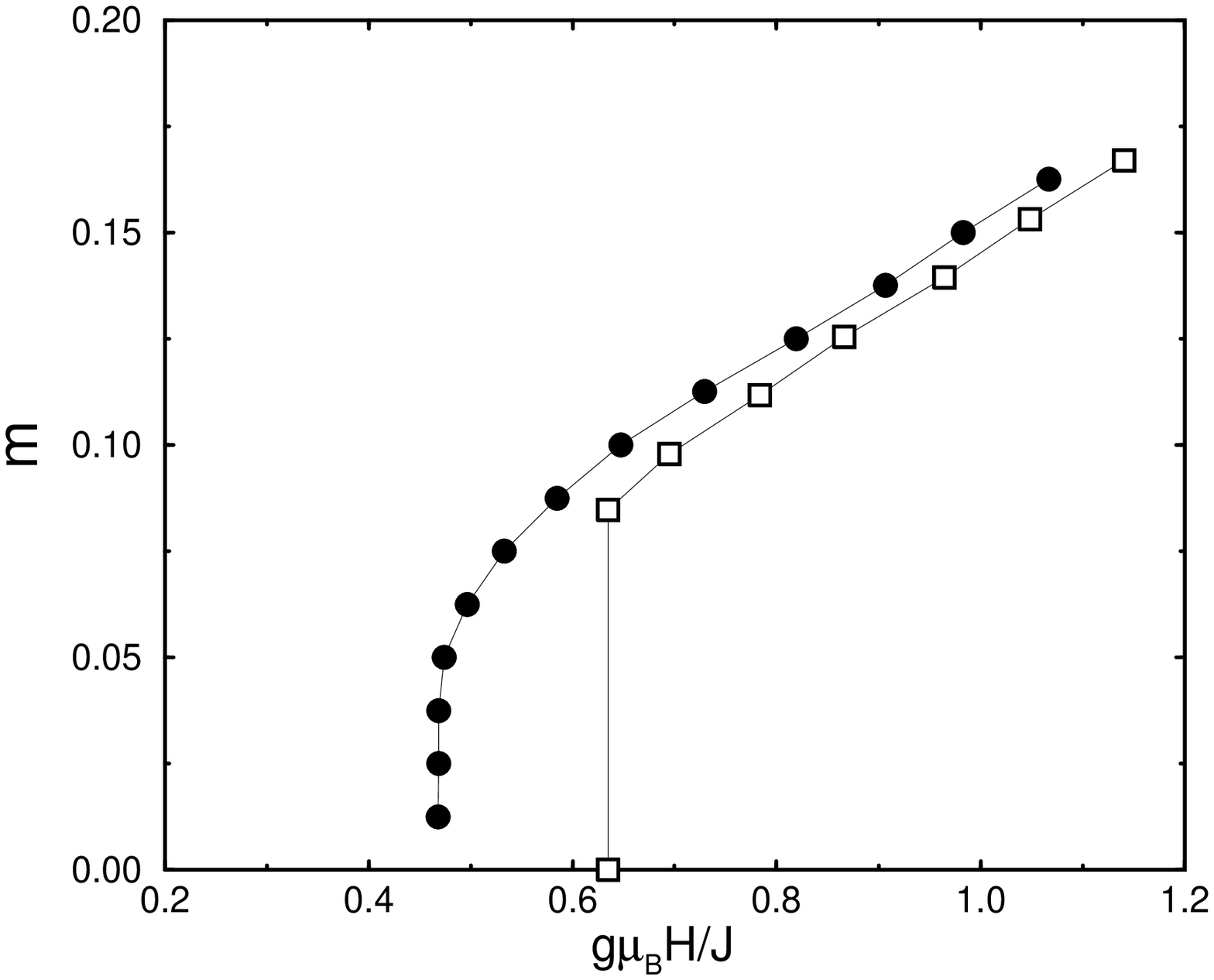}}
\caption[]{Filled circles: Magnetization for the adaptive modulation 
  ($K_0=1.7$, $\alpha = 0.35 $) as deduced from the corresponding
  curve in Fig.{\protect\ref{fig3}}.
  \newline
  Open squares: Magnetization with a dispersive elastic energy
  as discussed at the end of the paragraph. $\tilde K = 6 K_0$,
  $K_0$ and $\alpha $ unchanged.}	
\label{fig6}
\end{figure}

Within the self-consistent approach we calculate $E(m)$ per site as plotted in
Fig.\ref{fig3} (filled circles). $E(m)$ is convex but in contrast to the 
curves with fixed sinusoidal modulation it is continuous.
The convexity will be shown more clearly below in a
modified representation. Thus we have a continuous, {\em second order}
transition from the D phase to the adaptive I phase. The corresponding
magnetization $m(H)$ is shown in Fig.\ref{fig6} (filled circles).

 The enormous steepness of the continuous magnetization   
is explained by the following argument. For non-interacting 
spinons which are far enough from each other, the energy per site 
$E(m)-E(0)$ is proportional to the number of spinons and hence
to the magnetization $m$ (see also filled circles  in Fig.\ref{fig3} for
small $m$).  The proportionality constant $e_0$ is the energy of a
single spinon and determines the critical field $2e_0= g\mu_{\rm B}
H_c$, since two spinons are created by breaking one singlet.  
Because the spinons are exponentially localized (cf.
Fig.\ref{fig5}) two spinons at mutual distance $l$ have additionally
an exponential interaction $w (l) = w_0 \exp(-c l)$.  Here $c$ is a
constant of the order of the inverse correlation length and $w_0$ is a
proportionality constant which is positive for repulsion and negative
for attraction.  The typical distance of the spinons is $l=1/(2m)$
since each spinon carries spin $S=1/2$. Hence for not too large values
of $m$ the total energy in an external magnetic field $H$ equals
\setcounter{equation}{1}
\renewcommand{\theequation}{\arabic{equation}}
\begin{equation}
\label{energy}
E(m) - E(0) = g\mu_{\rm B}(H_c -H) m + w_0 2m
e^{-\frac{c}{2m}}\, .
\end{equation}
By minimizing this expression for repulsion ($w_0>0$) one derives
$H(m)$ which increases exponentially slowly just above $H_c$. This in
turn leads to the drastic increase of $m$ as depicted in
Fig.\ref{fig6}.

\begin{figure}
\begin{picture}(9,11)(0,0)
\put(2,10){$XY$}
\put(2,7){$XYZ$}
\put(0,4){\psfig{width=\columnwidth,file=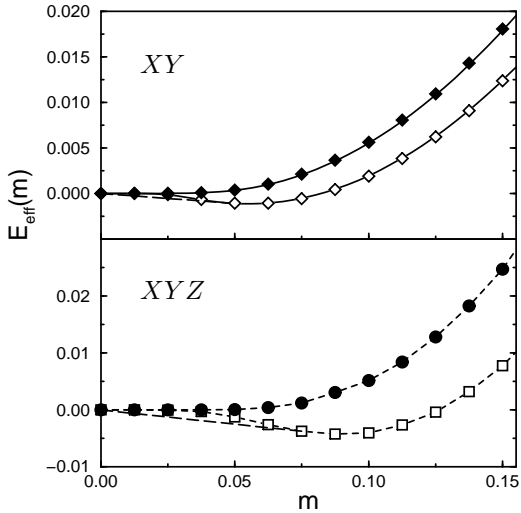}}
\put(-0.3,0){\caption[]{
  Affine representation of the ground state energy. The long-dashed
  lines indicate the convex hulls to the lower curves.
  $XY-$model: The upper solid curve shows $E_{\rm eff}$ for a 
  dispersionless elastic energy with $K_0=0.625$. 
  The lower solid curve shows $E_{\rm eff}$ for a dispersive
  elastic energy ($\tilde K=6K_0$) as discussed in the 
  following section. Both curves are obtained via 	
  the continued fraction technique. The filled and open diamonds 
  depict DMRG results for an 80 site chain.
  \newline
  $XYZ-$model: DMRG results in the dispersionless case (filled circles) and 
  for $\tilde K=6K_0$ (open squares) for $K_0=1.7$ and $\alpha = 0.35$.
  The dashed lines are guides to the eye.}
\label{fig7}}
\end{picture}
\end{figure}

To present the effects of soliton interaction more clearly we pass to
an affine representation of the ground state energy $E(m)$ by
investigating
\begin{equation}
\label{affin}
  E_{\rm eff}(m) := E(m) -  E(0) - g\mu_{\rm B}H_c m
\end{equation}
which would be constant zero if no interaction between the solitons
existed. Note that $E_{\rm eff}(m)$ is convex if and only if $E(m)$ is.
 In Fig.\ref{fig7} the generic resulting
curves are shown with filled symbols (solid line) for the $XY-$model 
and the spin isotropic $XYZ-$model. The results for the $XY-$model are 
obtained for an infinite system without frustration by a continued fraction technique
based on Green's functions \cite{uhrig98a}. Using this method the simpler 
solvability of the $XY-$model allows to iterate up to 80 times for an  
infinite chain with periodicities up to 120.
These data are included as an 
additional check that no spurious effects due to finite size or
insufficient iteration are investigated.

The results in Fig.\ref{fig7} for a dispersionless elastic energy comply
perfectly with exponentially repulsive solitons (\ref{energy}). 
There is no sign of a long range interaction $\propto 1/l$ as postulated
by Horovitz for finite cutoffs as they occur naturally in discrete
lattice models \cite{horov81a,horov87}.
In particular, no attraction for dispersionless elastic
energies are found \cite{note1}.

A dispersionless elastic energy is of course a drastic simplification of
the real phononic system.  Cross already argued \cite{cross79b} that a
pinning in $k$-space should influence the order of the transitions.
We expect that the D$\to$I phase transition becomes first order if the
elastic energy itself favors the distortion at $k=\pi$. This means
that $\hat{K}(\pi)$ is minimum if the elastic energy can be expressed as
$E_{\rm elast} =\frac{1}{2} \sum_k \hat{K}(k) |\delta_k|^2$.  The argument
compares for 
\noindent
a given wave vector $q$ close to $\pi$ the elastic energy
$\hat{K}(q)$ of a sinusoidal modulation (\ref{incomm}) with the one of an
array of domain walls with the same periodicity $2\pi/q$.  Since the
latter has also contributions of higher harmonics $\pm3q$, $\pm5q$, $\pm7q$, ...
its elastic energy is $\sum_{n} |a_{(2n+1)q}|^2 \hat{K}(q)$ where the 
coefficients $|a_{(2n+1)q}|^2$ are symmetric about $\pi$. Thus the  
elastic energy is higher than the one
for the sinusoidal modulation. By this mechanism higher harmonics 
are suppressed due to the elastic energy leading to a smoother 
and more sinusoidal modulation.
If the convex curve for the adaptive modulation in Fig.\ref{fig3} is
influenced in a way to approach the discontinuous curve for sinusoidal
modulation one must expect a region of concavity for small $m$. Hence
the convex hull differs from the curve itself and a jump in the
magnetization occurs: the transition is first order.  Put differently,
we expect that a dip in the elastic energy at the zone boundary 
leads to an {\em attraction} of the solitons.

To investigate our hypothesis numerically we use $\hat{K}(k) = K +2\tilde
K\cos(k)$ with $K_0= K-2\tilde K$ kept fixed to refer to the same
amplitudes in the D phase.  This elastic energy corresponds in real
space to
\begin{eqnarray}
  E_{\rm elast}&=&\frac{1}{2} \sum_i \left(K\delta_i^2 +2\tilde K
    \delta_i \delta_{i+1} \right) \nonumber
\\
 &=&\frac{1}{2} {\mbox{\boldmath
      $\delta$}}^+ \, {\bf K}\, {\mbox{\boldmath $\delta$}} \ ,
\end{eqnarray}
where {\mbox{\boldmath $\delta$}} is a vector with components
$\delta_i$ and ${\bf K}$ is a cyclic tridiagonal $L\times L$
symmetric matrix of coupling constants with diagonal elements $K$ and
off-diagonal elements $\tilde K$.  Generic results for the energies
$\tilde E(m)$ in affine representation are depicted with open symbols
(solid line) in Fig.\ref{fig7}. We find indeed a concavity for small 
values of $m$. This implies soliton attraction and a first order 
transition.

Furthermore, we show in Fig.\ref{fig6} the resulting magnetization
curves with and without dispersion of the elastic energy.  The
difference between the second order transition for the elastic energy
without dispersion and the first order transition with dispersion is
clearly visible. Additionally, the critical field $H_c$ at which the
transition occurs rises on inclusion of the dispersion.  This complies
also with the above consideration since the energy of a single soliton
rises due to $K+2\tilde K \cos(k) > K_0$ except at $k=\pi$ for 
$\tilde K>0$.  

\begin{figure}
\begin{picture}(9,6)(0,0)
\put(0,4){$\delta_i$} \put(4.5,1.2){\it i}
\put(0,1){\psfig{width=\columnwidth,file=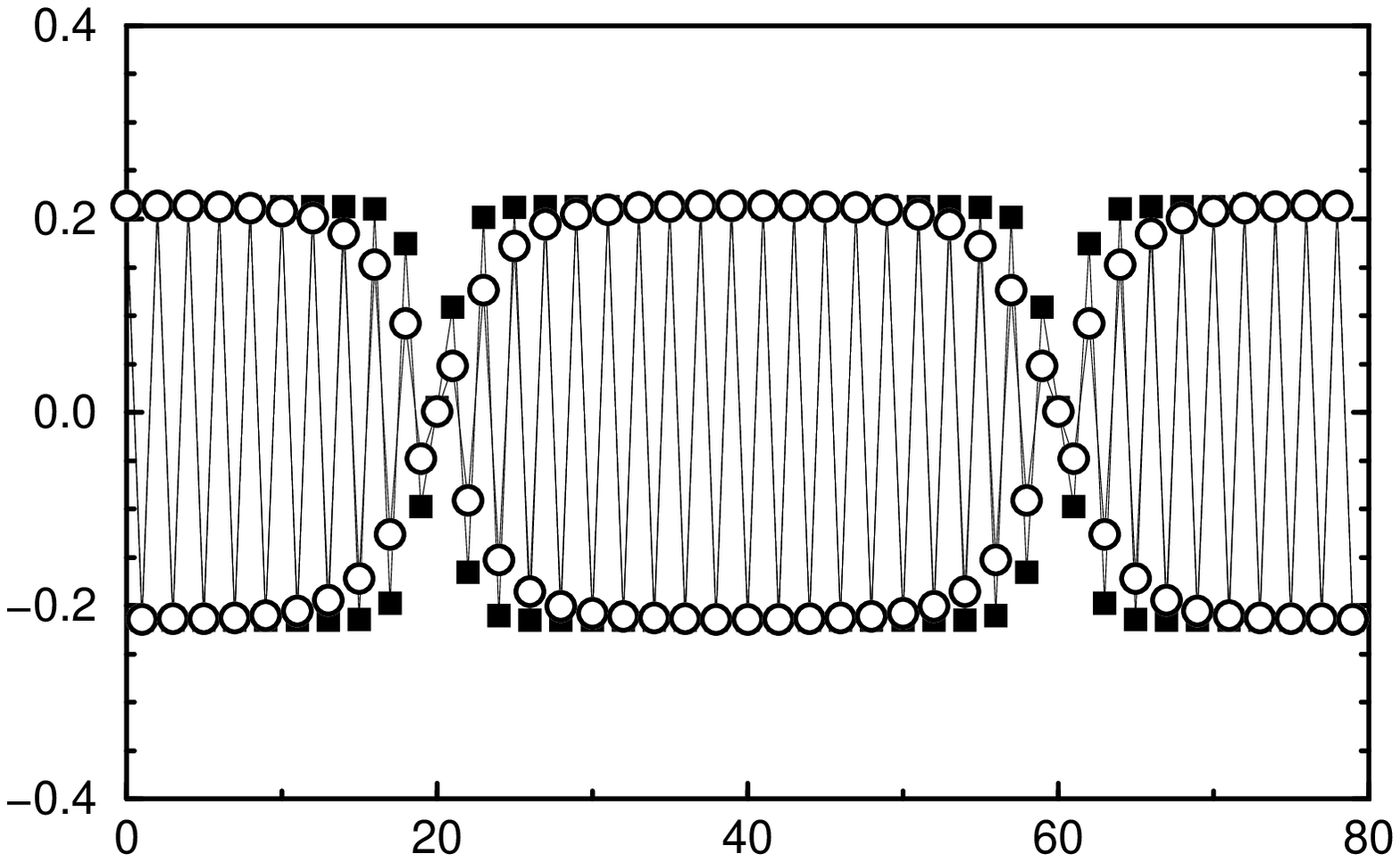}}
\put(-0.3,0){\caption[]{Modulations for $S_z = 1$ for the  same 
  parameters as in Fig.\protect{\ref{fig7}} for the $XYZ-$model.}
\label{fig9}}
\end{picture}
\end{figure}

Finally, in Fig.\ref{fig9} the modulation patterns with and
without dispersion are compared. Indeed, the inclusion of $\tilde K>0$
makes the modulation softer and more sinusoidal.  In conclusion 
our numerical results convincingly corroborate our expectations for the 
effect of a dispersive elastic energy.

Numerically, we are not able to decide whether an arbitrarily small
$\tilde K$ already yields a soliton attraction. For smaller values of $\tilde
K$ the minima in the affine representation occur for smaller and
smaller magnetization and they are more and more shallow. We expect
that the soliton attraction exists down to arbitrarily small values of
$\tilde K$ but it may become irrelevant in practice due to the
exponential smallness of the corresponding energies.

We also investigated negative values of $\tilde K$. No qualitative
change of the soliton interaction was found in comparison to the 
dispersionless case. The iterative procedure,
however, becomes quite unstable already for small negative values of
$\tilde K$.

\subsection{Adiabatic Gaps}

So far we aimed at the average magnetization as a function of the
applied magnetic field. Another interesting quantity which is
accessible once $E(m)$ can be computed are the adiabatic gaps.
It is a so far unsettled question whether spin-Peierls systems have
or have not gaps in the incommensurate phase.

On the one hand, it seems clear that the incommensurate modulation pattern
can be shifted along the chains without energy cost.  This is
certainly true in the con\-ti\-nu\-um description and thus most probable
also for not too small correlation lengths.  This quasi-continuous
symmetry gives rise to quasi-Goldstone bosons called phasons which are
gapless \cite{bhatt98}. They do not change the spin sector and thus have
 $\Delta S_z =0$.  The physics of phasons is beyond an adiabatic
treatment of the lattice distortion since within an adiabatic
treatment  the distortion is assumed to be fixed.

A different issue is the question whether the gaps $\Delta_{\pm}$
corresponding to $\Delta S_z =\pm1$ are finite or not.  Note that these
gaps do not need to be equal since the spin rotation symmetry is broken for
finite magnetization.  From a non-adiabatic viewpoint one can still
infer from the smoothness of the $E(m)$ curves that there are
no such gaps in the I phase since the modulation adapts always to the
average magnetization. Applying, however, an operator like $S^+(k)$ or
$S^-(k)$ \cite{mulle81,uhrig98a} and asking for the accessible
excitation spectrum may lead to a different answer.  These operators
act only on the spin part of the ground state and leave the modulation
unchanged. Thus it is not unreasonable to expect that the gapless
excitations are not accessible if their access required a
re-arrangement of the whole, in reality three-dimensional, modulation.
The underlying question is whether the states $S^{\pm}(k)|S_z\rangle$
are orthogonal to $|S_z\pm 1\rangle$ or not, if we denote by
$|S_z\rangle$ the ground state for the magnetization $S_z$.

 Here we
will investigate the simpler question whether in the strictly
adiabatic framework the gaps $\Delta_{+\-}$ are finite or not.  Uhrig
{\it et al}. \cite{uhrig98a} were only able to compute $\Delta_+ +\Delta_-$ 
since this quantity did not require the knowledge of the corresponding magnetic
field.
 
We define by $E(m,H):= E(m) - mg\mu_{\rm B} H$ the
ground state energy with self-consistently optimized modulation
$\{\delta_i\}$.  By
\begin{eqnarray} \nonumber
  E_\pm(m,H) &:=& \frac{1}{L}\langle \hat H_{\rm chain} \rangle\Big|_{S_z =
    mL\pm1} 
\\
&&+ \frac{K_0}{2L}\sum_i \delta_i^2 -
  \left(m\pm\frac{1}{L}\right) g\mu_{\rm B} H
\end{eqnarray}
we denote the ground state energy with one additional spin flipped
upward (+) or downward (-), respectively, {\em but} with the
modulation $\{\delta_i\}$ belonging to $S_z = mL$, not to $S_z =
mL\pm1$.  This means that for $E_\pm(m,H)$ the modulation is
not optimized for the given magnetization. This corresponds to the
situation accessible by application of $S^+(k)$ or $S^-(k)$ without
reaction of the lattice part.  Then the gaps are defined by
\begin{equation}
  \Delta_\pm(m) = E_\pm(m,H) - E(m,H) \, .
\end{equation}

\begin{figure}
\begin{picture}(9,5.5)(0,0)
\put(0,0.5){\psfig{width=\columnwidth,file=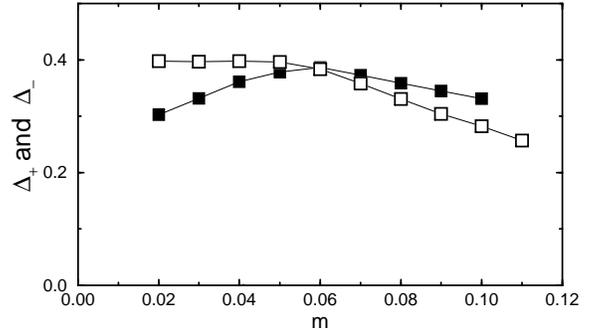}}
  \put(-0.3,0){\caption[]{The energy gaps $\Delta_+$ (filled squares)
      and $\Delta_-$(open squares) as a function of the magnetization
      for $L=100$, $\alpha = 0.35 $ and $K_0 = 2.38$}
\label{fig10}}
\end{picture}
\end{figure}

The gaps $\Delta_+$ and $\Delta_-$ for a 100 site ring are displayed in
Fig.\ref{fig10} for $\alpha = 0.35 $ and $K_0 = 2.38$ corresponding to
$\delta \approx 0.14$ in the D phase.  
Finite size effects are not yet completely negligible, but the
qualitative behavior is the one shown and is in agreement with
previous self-consistent renormalized Hartree-Fock results
\cite{uhrig98a}.  \unitlength1cm
Most importantly, we can show by Fig.\ref{fig10} that both gaps are
indeed finite and of equal order of magnitude.  It is interesting that
apparently $\Delta_+$ is smaller at small magnetization and $\Delta_-$
is smaller at larger magnetization.  At least in the adiabatic
approach, we can show that even the I phase has gaps. It would be
interesting if any experimental evidence in favor of the existence
of these gaps was found.

\section{Computational details}
In our DMRG calculation
\cite{white92,white93} we apply periodic boundary conditions to
minimize finite size effects. We keep 128 (64) states in the
truncation procedure. To account for the incommensurate structure we
use the finite size algorithm \cite{white92,white93}.  In the first
steps where the system is iteratively increased we use the reflection
of the left hand block to build up the superblock although the
reflection symmetry is not given at this stage. This initial error is
reduced either by supplementary sweeps through the system of the
desired length or by intermediate sweeps through the system the length
of which is commensurable with the lattice modulation.  We tested the
accuracy of the DMRG results by comparing the lowest energies of the
$XY$-Model with sinusoidal modulation in different $S_z$-subsectors 
with exact results for a 60-site ring.
Keeping 128 (64) states we find the typical error to be smaller than
$10^{-6}$ ($10^{-5}$) for $S_z=0$ and $10^{-5}$ ($10^{-4}$) in
higher $S_z$-subsectors. 
For the selfconsistent calculations in the case of adaptive modulations
we used the sinusoidal modulation (\ref{incomm}) as a starting configuration
for the curves presented. For larger dimerizations, however, 
it is more convenient to start with a step-like modulation since the 
correlation lengths become very small. Having calculated the ground state 
in the corresponding  $S_z$-subsector we use equation (\ref{deltai}) 
to deduce the improved set of $\{\delta_i\}$. This step is repeated 
(typically 6 to 10 times) until the change of the ground state energy becomes 
sufficiently small, i.e. of the order of the truncation error. 

For the $XY$-Model we can compare the selfconsistent DMRG results 
for finite chains with these of the continued fraction technique 
\cite{uhrig98a} in the thermodynamic limit.  We find that the error 
due to finite size effects and due to the truncation of the Hilbert space
is at most of the order of $10^{-4}$ (see upper part of Fig.\ref{fig7}). 
This accuracy is by far sufficient for the presented qualitatively 
analysis of the phase transition.

\section{Summary and Discussion}

In this work we considered modulated $S=\frac{1}{2}$ Heisenberg
chains with finite magnetization. Three classes of modulation
were investigated: (i) fixed dimerization, (ii) fixed 
incommensurate sinusoidal modulation and (iii) adaptive incommensurate
modulation. Our main interest was to investigate the dependence of the 
magnetization $m$ on the applied magnetic field $H$.

For scenario (i) we found a second order transition to finite magnetization 
by means of the finite size DMRG method in agreement with previous calculations. 
The increase of $m$ just above the critical field $H_c$ is characterized by
a square root behavior $m\propto \sqrt{H-H_c}$.

For scenario (ii) we showed that the system favors an incommensurability
corresponding to the magnetization $q=2k_{\rm F} = 2\pi (1/2+m)$. 
A prominent first order transition
is found. This finding could be explained by computing the discontinuous 
dependence of the ground state energy $E(m)$ on $m$ at $m=0$.
The discontinuity is linked to the discontinuous jump of the
root-mean-square of the local distortions on passing from
dimerization ($q=\pi$) to a long-wave length modulated 
dimerization ($q\approx \pi$).

In scenario (iii) we determined iteratively the modulation which 
minimizes the total energy including a quadratic elastic energy.
Again we find a periodicity corresponding to the magnetization
$q = 2\pi (1/2+m)$. The modulation, however, corresponds for low
magnetization rather to a soliton lattice. This means one has
differently dimerized regions separated by domain walls. Each 
domain walls carries one $S=1/2$. We find a crossover from the
solitonic picture at low magnetization to a sinusoidal modulation
at higher magnetizations.
 The transition
to finite magnetization is second order although the increase is
exponentially steep. The inclusion of a positive dispersion of 
the elastic energy alters the order of the transition. It is 
first order then. An exponential attraction of the solitons was 
identified.

As another interesting quantity we calculated the adiabatic
gaps $\Delta_{+/-}$ corresponding to the increment (decrement)
of the magnetization by unity. The independent determination of these
gaps requires the complete knowledge of $E(m)$.
The calculation was also done
for scenario (iii). It was shown that these gaps are finite
in the adiabatic treatment. 

The second order phase transition in the commensurate case
(\ref{comm}) is in agreement with the fact that measurements under
applied field for instance on Cu$_2$(C$_5$H$_{12}$N$_2$)$_2$Cl$_4$
show no hysteresis effects \cite{chabo97}. This substance is found 
to be an antiferromagnetic Heisenberg ladder which is equivalent to 
a strongly dimerized quasi-one-dimensional Heisenberg chain. 
The magnetization increases continuously. The
expected square root behavior near $H_c$, however, was not observed.

We do not know of a substance which can be described by  
pure sinusoidally modulated exchange couplings. The modulation
in the incommensurate phase of CuGeO$_3$ is in fact very
close to a sinusoidal modulation \cite{kiryu95,kiryu96a}.
Recently Lorenz {\it et al.}\cite{loren98}
measured the magnetic field dependence of 
the spontaneous strain $\epsilon(H)$ in CuGeO$_{3}$ which is
in first approximation proportional to the elastic energy
associated to the lattice distortion. It
decreases very fast near $H_c$ and saturates approximately at 1/4 of
the  value in the dimerized phase for $H\approx 22$T.
One can conclude that 
there is a crossover from a solitonic distortion pattern
for small magnetizations to a sinusoidal one for larger
magnetizations.
Our model allows - for  parameter values reasonable for CuGeO$_3$
within a one-dimensional approach $\alpha=0.35$, $K \approx 18$ 
($\Rightarrow \delta=0.014$ in the D phase \cite{bouze97a,riera95,fabri98})
 -  to describe the above crossover quantitatively \cite{loren98}.

The feature so far not understood in CuGeO$_{3}$ is the first
order phase transition D$\to$I. From our findings it is tempting
to attribute this weak first order property to a positive dispersive
elastic energy, i.e.\ a dip in $\omega(k)$ at $k=\pi$.
Unfortunately, there is no experimental indication for such a feature 
in the phonon spectra \cite{brade98}.
The spring constant $\hat{K}(k)$, however, in the adiabatic treatment is 
proportional to $\omega(k)/g^2(k)$ where $\omega(k)$ is the phonon
energy
measured by Braden {\it et al.} \cite{brade98}. The 
momentum-dependent spin-phonon coupling
$g^2(k)$ is not known presently and may account for the dispersive
behavior needed to explain the first order transition D$\to$I observed in
CuGeO$_{3}$.
At the present stage, we may also speculate  that the neglected
interchain couplings \cite{uhrig97a} are decisive for the order
of the transition. From our present results we understand that
the order of the transition is influenced by the microscopic details
of the model.

The finding of finite adiabatic gaps in the incommensurate phase
should encourage experimental work to verify or to falsify
this feature, for instance, in CuGeO$_{3}$.

\section*{Acknowledgement}

We thank B. B\"uchner, A.P. Kampf, Th. Lorenz, J.P. Boucher and Th. Nattermann
for numerous discussions. One of us (GSU) likes to thank the hospitality
of the NHMFL, Tallahassee.

This work was supported by the Deutsche Forschungsgesellschaft with in
Sonderforschungsbereich 341.


\end{document}